\documentclass[runningheads]{llncs}
\usepackage{graphicx}
\usepackage{amsmath}
\usepackage{multirow}
\usepackage{makecell}
\begin{document}
\title{Shifted Chunk Encoder for Transformer Based Streaming End-to-End ASR\thanks{Supported by the National Innovation 2030 Major
S\&T Project of China under Grant 2020AAA0104202.}}
\author{Fangyuan Wang\inst{1}\orcidID{0000-0002-6482-4522} \and
Bo Xu\inst{1,2,3}}
\institute{Institute of Automation, Chinese Academy of Science, Beijing, China \and
School of Future Technology, University of Chinese Academy of Sciences
, Beijing, China \and
School of Artificial Intelligence, University of Chinese Academy of Sciences, Beijing, China\\
\email{\{fangyuan.wang,xubo\}@ia.ac.cn}}
\maketitle
\begin{abstract}
Currently, there are mainly three kinds of Transformer encoder based streaming End to End (E2E) Automatic Speech Recognition (ASR) approaches, namely time-restricted methods, chunk-wise methods, and memory-based methods. 
  Generally, all of them have limitations in aspects of linear computational complexity, global context modeling, and parallel training.
  In this work, we aim to build a model to take all these three advantages for streaming Transformer ASR.
  Particularly, we propose a shifted chunk mechanism for the chunk-wise Transformer which provides cross-chunk connections between chunks. Therefore, the global context modeling ability of chunk-wise models can be significantly enhanced while all the original merits inherited.
  We integrate this scheme with the chunk-wise Transformer and Conformer, and identify them as SChunk-Transformer and SChunk-Conformer, respectively.
  Experiments on AISHELL-1 show that the SChunk-Transformer and SChunk-Conformer can respectively achieve CER 6.43\% and 5.77\%.
  And the linear complexity makes them possible to train with large batches and infer more efficiently.
  Our models can significantly outperform their conventional chunk-wise counterparts, while being competitive, with only 0.22 absolute CER drop, when compared with U2 which has quadratic complexity.
  A better CER can be achieved if compared with existing chunk-wise or memory-based methods, such as HS-DACS and MMA. Code is released.\footnote{https://github.com/wangfangyuan/SChunk-Encoder.}.

\end{abstract}

\section{Introduction}

In the past decades, 
ASR with E2E models has achieved great progress, and has become a popular alternative to the hybrid ASR models equipped with conventional Hidden Markov Model (HMM)/Deep Neural Network (DNN). Currently, Connectionist Temporal Classiﬁcation (CTC)\cite{Li18,Graves06}, Recurrent Neural Network Transducer (RNN-T)\cite{Battenberg17}, and Attention based Encoder-Decoder (AED)\cite{chan16,Gulati20}
are the three mainstream E2E systems.
Also, efforts to conduct performance comparisons\cite{Prabhavalkar17} or the combination\cite{Watanabe17,Miao20} of these models have been made.
Recently, Transformer\cite{Vaswani17} has become a prevalent architecture, outperforming RNN\cite{yuanyuanzhao} in AED systems\cite{Prabhavalkar17}.  
Furthermore, 
Transformer can also use as an encoder with CTC\cite{Li18} or Transducer\cite{Chen21}.
And very recently, the Conformer\cite{Gulati20} has been proposed which augments Transformer with convolution neural networks (CNN). 
Both Espnet\cite{Guo21_espnet} and WeNet\cite{Yao21_wenet} have shown that Conformer can bring significantly performance gains on a wide range of ASR corpora.

\begin{figure}
\centering
\includegraphics[width=7cm]{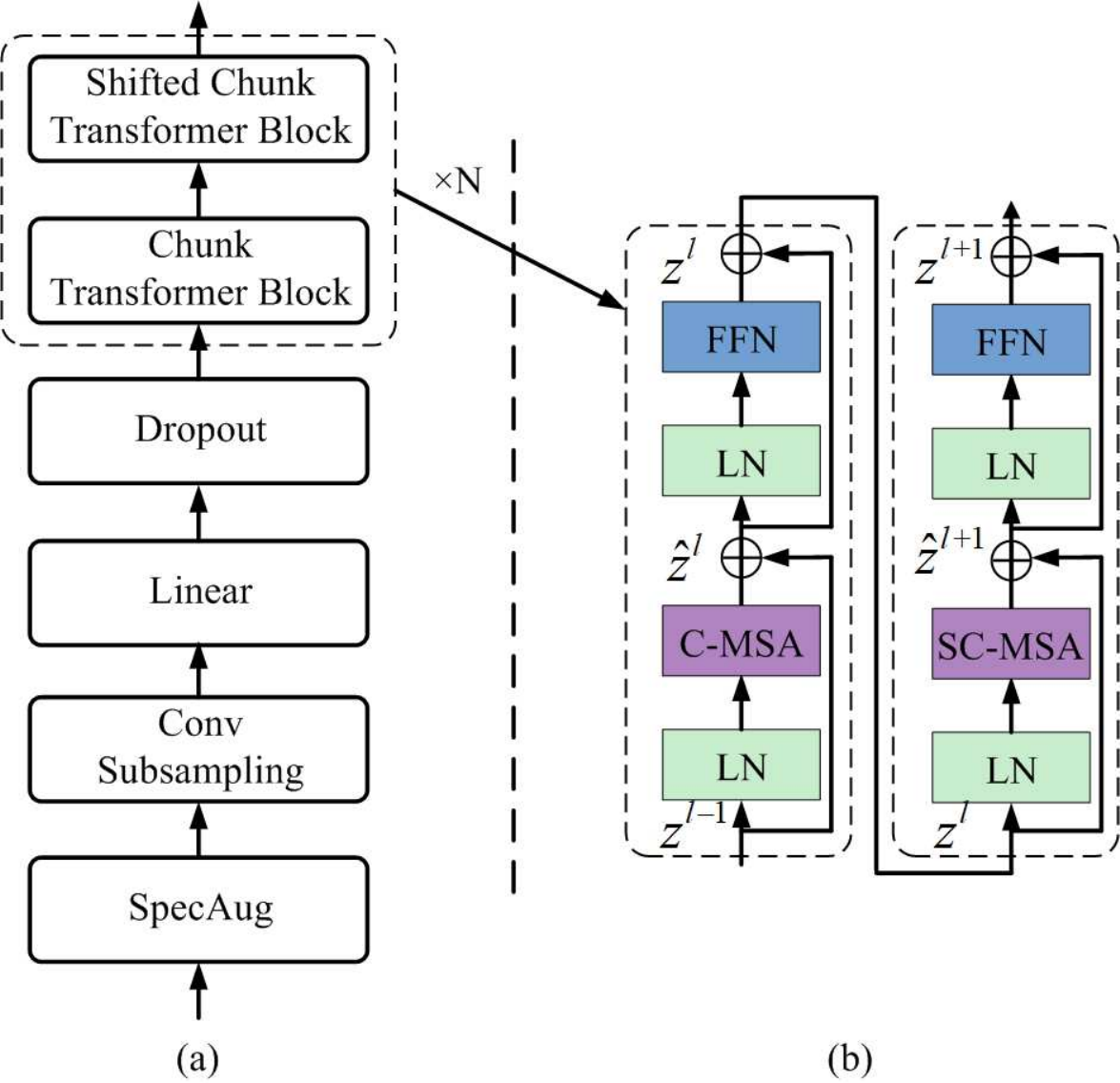}
\caption{(a) The architecture of SChunk-Transformer, \emph{N} is set to 6 by default; (b) two successive blocks (notation presented with
Eq. (3)).}
\label{fig:architecture}
\end{figure}

The great success of Transformer and its variants urge people to explore its adaption for streaming ASR. 
However, two issues make vanilla
models impractical for streaming ASR.
First, the calculation of self-attention depends on the entire input sequence.
Second, the computation and memory usage grow quadratically to the length of the input sequence.
Actually, several methods have been proposed to alleviate these issues.
1) Time-restricted methods\cite{Yu20_arxiv,Tripathi20_arxiv,zhang2020unified_arxiv,wu2021u2++_arxiv,WNARS_arxiv,CUSIDE_arxiv} where the attention computation only uses past input vectors and limited future inputs. However, the time and memory complexities of these methods are still quadratic, which may introduce a significant latency for long inputs.  
2)Chunk-wise methods\cite{Chen21,Tian2020_icassp,Li2021_head_icassp} typically
evenly partition the input into chunks and then calculate attention only within these chunks as monotonic chunk-wise attention (MoChA)\cite{Chiu2018_iclr}. 
They have linear complexities but usually suffer dramatic performance drops as the reception field of attention is limited within local chunks. 
3)Memory-based methods\cite{Zhang2020_interspeech,Inaguma12020_interspeech,Shi2021_emformer_icassp} utilize the solution of chunk-wise methods to reduce running time while employing an auxiliary contextual vector to memorize the history information.
However, these vectors break the parallel nature of Transformer, typically requiring a longer training time.

In this paper, we aim to build a streaming Transformer which can compute in linear complexity, capture global history context and parallel train simultaneously.
Under the guidance of this goal, we find inspiration from Swin Transformer\cite{liu2021Swin} and 
introduce the idea of shifted windows into streaming ASR. 
In detail, we propose a shifted chunk mechanism for chunk-wise Transformer models. This mechanism allows the computation of attention to cross the boundary of chunks, thus can significantly enhance the model power, while keeping linear complexity and parallel training.
We integrate the proposed mechanism into Transformer and Conformer and get SChunk-Transofromer and SChunk-Conformer, respectively. 
And we have conducted ablation studies and comparison experiments on
AISHELL-1\cite{Bu2017Aishell-1}. The results show that Schunk-Transformer
and Schunk-Conformer can respectively achieve CER 6.43\%
and 5.77\% when set the chunk size to 16, which significantly
surpass their conventional chunk-wise counterparts. 
When compared with U2\cite{zhang2020unified_arxiv}, which is a strong baseline model using the time-restricted method, our models can still be competitive with only an absolute 0.22 CER drop for SChunk-Conformer but be more efficient to train and infer. Superior performance can achieve if compared with other existing chunk-wise or memory-based methods, such as HS-DACS\cite{Li2021_head_icassp} and MMA\cite{Inaguma12020_interspeech}.

\section{Shifted Chunk Encoder}

For convenience, we take SChunk-Transformer as an illustrative encoder to describe the mechanism of the shifted chunk. 

\subsection{Overall Architecture}
As illustrated in Fig.~\ref{fig:architecture}(a), our proposed encoder first processes the input audios with SpecAug\cite{ParkSpecAug19}, convolution subsampling, and other frontend layers as conventional Transformer ASR, and then with several consecutive chunk Transformer blocks and shifted chunk Transformer blocks. The distinctive feature of our model is the use of chunk Transformer block and successively shifted chunk Transformer block to replace chunk Transformer blocks.

\subsection{Shifted Chunk Transformer Block}
We build the SChunk-Transformer block by replacing the multi-head self attention (MSA) in a Transformer block with a module based on shifted chunks (described in Section II.C), with other layers kept the same, see Fig.~\ref{fig:architecture}(b).
The SChunk-Transformer block is composed of a shifted chunk based MSA
module, followed by a 2-layer Feed Forward Network (FFN)
with GELU nonlinearity in between. It applies a LayerNorm (LN) layer
before each MSA and FFN module and adds a residual
connection after each module.

\subsection{Shifted Chunk Based Self-Attention}
\subsubsection{Chunk-wise Self-attention}
The vanilla Transformer\cite{Vaswani17} uses global MSA to compute the dependencies between a frame and all the other frames. To be efficient, we calculate self-attention within evenly partitioned
non-overlapped chunks.
If an audio of \emph{L} frames and each chunk has \emph{W} frames, the complexities of computing a global MSA and a chunk based MSA are\footnote{We omit softmax computation in determining complexity}:

\begin{equation}
  \Omega(MSA) = 4L\cdot{C^{2}}+2L^{2}\cdot{C}
  \label{eq1}
\end{equation}
\begin{equation}
  \Omega(C\text{-}MSA) = 4L\cdot{C^{2}}+2N\cdot{L\cdot{C}}
  \label{eq2}
\end{equation}
where \emph{C} is the feature dimension, the former is quadratic to \emph{L}, and the
latter is linear when \emph{W} is a fixed value. 
Global MSA is generally unaffordable for a large \emph{L}, which may introduce a significant latency for time-restricted methods.

\begin{figure}[t]
  \centering
  \includegraphics[width=8cm]{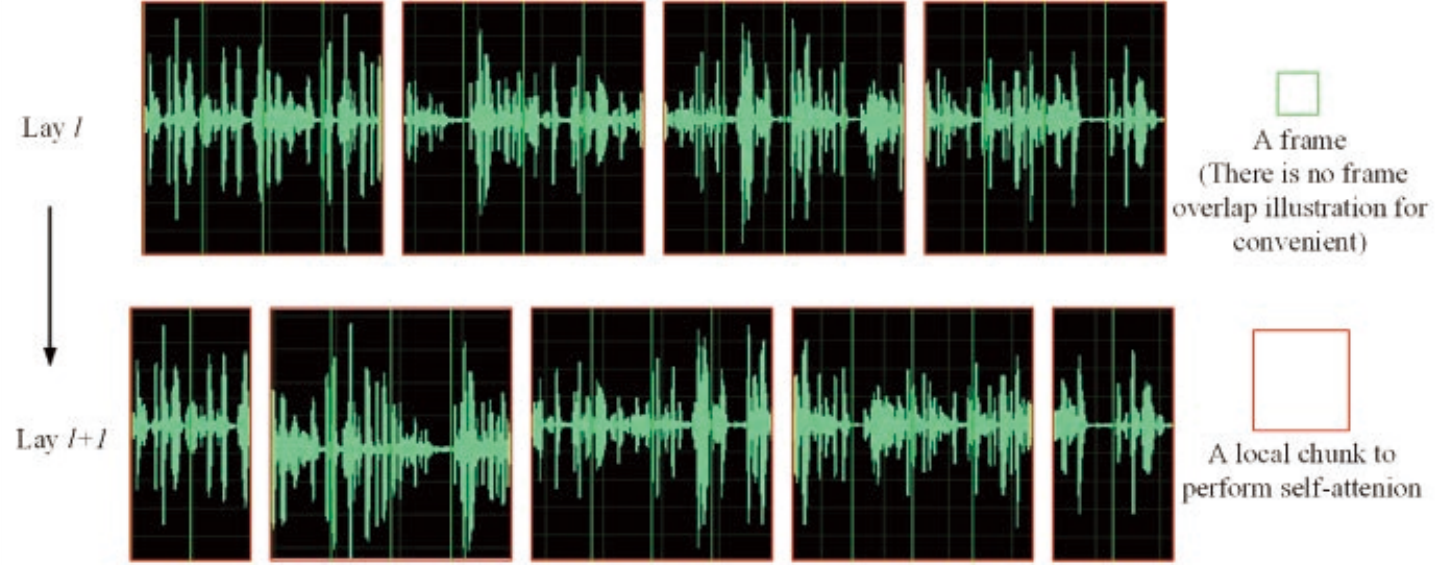}
  \caption{ An illustration of the shifted chunk approach for computing self-attention. In layer \emph{l} (top), 
  self-attention is computed in local chunks which are got by a regular chunk partitioning scheme. In the next layer \emph{l+1} (bottom), the self-attention computations are conducted in new chunks which cross the previous chunks in layer l and got by shifting.}
  \label{fig:shifted_chunks}
\end{figure}

\subsubsection{Shifted Chunk Partitioning in Successive Blocks}
The chunk based MSA lacks connections across chunks, which limits its modeling power. 
We propose the shifted chunk partition approach to introduce cross-chunk connections while maintaining the efficiency of chunk-wise computation.
As shown in Fig.~\ref{fig:shifted_chunks}, 
we use the regular partitioned chunks
followed by the shifted partitioned chunks consecutively.
The regular chunk partitioning strategy starts from the audio, 
and the feature sequence of 16 frames is evenly partitioned into 4 chunks of size 4 (\emph{W}=4).
Then, the shifted partition is shifted from the preceding layer, by displacing the chunks by $\lfloor{\emph{W}/2}\rfloor$
frames from the regularly partitioned chunks.

With the shifted chunk partitioning approach, the Chunk-Transformer block and SChunk-Transformer block are computed as:

\begin{equation}
	\begin{aligned}	 
    \hat{z}^{l}&=C\text{-}{MSA}(LN(z^{l-1}))+z^{l-1}\text{,}\\
    z^{l} &=FFN(LN(\hat{z}^{l}))+\hat{z^{l}}\text{,}\\
    \hat{z}^{l+1} &=SC\text{-}{MSA}(LN(z^{l}))+z^{l}\text{,}\\
    z^{l+1} &=FFN(LN(\hat{z}^{l+1}))+\hat{z}^{l+1}\\
    \end{aligned}\label{eq3}
\end{equation}
where $\hat{z}^{l}$ and $z^{l}$ denote the outputs of the (S)C-MSA and the FFN for block \emph{l}, respectively;
S-MSA and SC-MSA denote chunk based multi-head self attention using regular and shifted chunk partitioning configurations, respectively.

\begin{figure}[t]
  \centering
  \includegraphics[width=10cm]{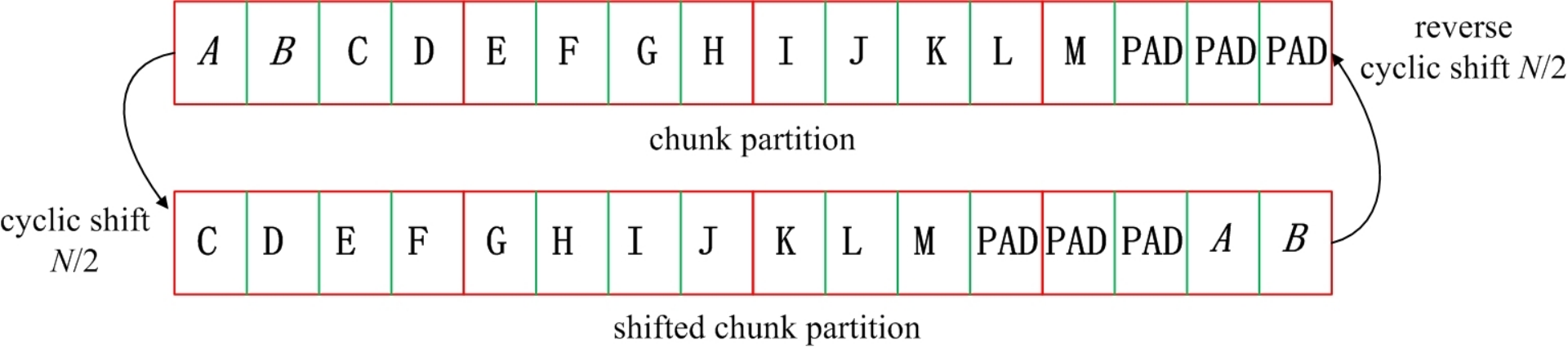}
  \caption{An illustration of efficient batch computation for self-attention in shifted chunk partitioning.}
  \label{fig:cyclic_shift}
\end{figure}

\subsubsection{Efficient Batch Computation for Shifted Chunks}
The first issue of shifted chunk partitioning for batch computation is the
the difference in audio lengths. 
To be evenly partitioned, we pad audios in a batch to the same length, which is a little longer than the longest one in the batch while can be evenly divided by the chunk size. 
Another issue is that shifted chunk partitioning will result in more chunks, and some chunks will be smaller than \emph{W}, see Fig.~\ref{fig:shifted_chunks}.
We use a batch computation approach by cyclic-shifting the regular partitioned chunks from head to tail to get the shifted partitioned chunks, and reverse cyclic-shifting the shifted partitioned chunks from tail to head to re-get the regular partitioned chunks, see Fig.~\ref{fig:cyclic_shift}. 
With the cyclic-shift, the number of batched chunks remains the same as that of regular chunk partitioning, and thus is also efficient. 

\begin{figure}[t]
  \centering
  \includegraphics[width=9cm]{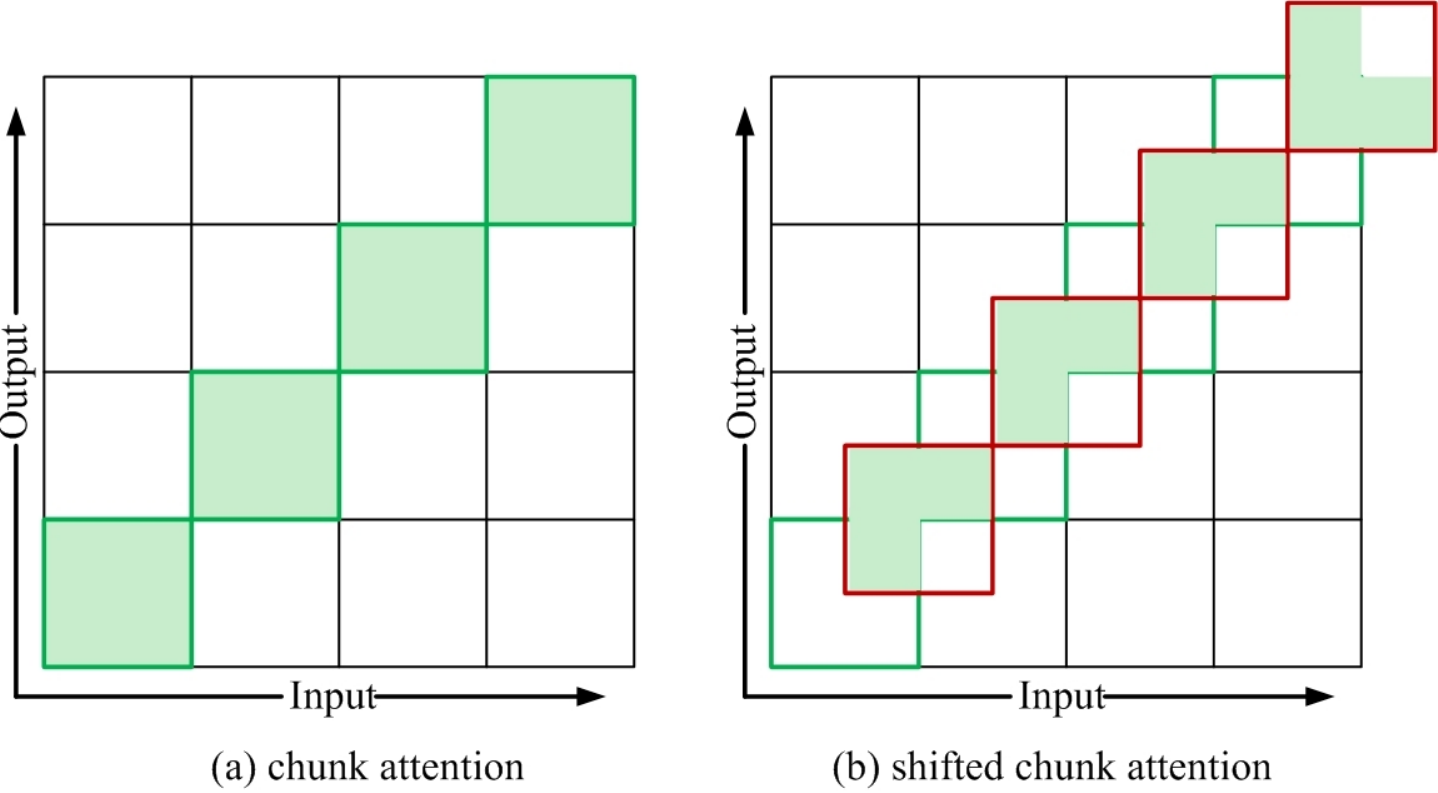}
  \caption{Illustration of masks for chunk attention and shifted chunk attention.}
  \label{fig:chunk_attention_mask}
\end{figure}

\subsubsection{Shifted Chunk Attention Mask}
As shown in Fig.~\ref{fig:chunk_attention_mask}(a), the chunk based self-attention can compute using a chunk-wise attention mask to support streaming.
However, for the shifted chunks, we need to mask out some areas as shown in Fig.~\ref{fig:chunk_attention_mask}(b) to make sure frames can only attend to their preceding ones when calculating the chunk-wise attention of SC-MSA.

\section{Streaming ASR with Shifted Chunks }
\subsection{Streaming Encoder and Decoder}
The SChunk-Transformer equipped with an attention mask can also support the streaming process as other chunk-wise methods. 
The casual convolution is used in SChunk-Conformer to make the CNN modules support streaming as in \cite{zhang2020unified_arxiv}.

We generally follow the decoder of U2\cite{zhang2020unified_arxiv} that uses a hybrid CTC/Attention decoder.
The CTC decoder outputs the first pass hypotheses in a streaming way. And then, the Attention decoder outputs the final results using full context to rescore the first pass hypotheses. 

\subsection{Streaming Inference}
In the inference stage, the encoder consumes the inputs chunk by chunk.
There is no shift in SC-MSA for the first chunk, and it degrades to behavior as C-MSA in this case without any impact on the first word prediction.
For the subsequent chunks, we need to cache the past chunks and concatenate them with the current chunk as the input for the encoder, like    the time-restricted methods.
Once the CTC decoder receives the output of the encoder, it generates output immediately.
At the end of an utterance, the Attention decoder is triggered to re-score the output of the CTC decoder to get a better utterance level result.

\section{Experiments}
\subsection{Data}
We evaluate the proposed models on AISHELL-1\cite{Bu2017Aishell-1},
which contains 150 hours of the training set, 10 hours of dev set and 5
hours test set, the test set consists of 7176 utterances in total.
The official vocabulary contains 4233 tokens.

\subsection{Experimental Setup}
We implement models using the WeNet toolkit\cite{Yao21_wenet} and verify on two NVIDIA Gefore RTX 3090 GPUs (24G).
For most hyper-parameters, we follow the recipes of WeNet.
(FBank) splice 3-dimensional pitch computed on 25ms window with 10ms shift as input feature. And speed perturbation with
0.9, 1.0, and 1.1 are done to get 3-fold data.
SpecAug\cite{ParkSpecAug19} is applied with 2 frequency masks with a maximum frequency mask (F = 50), and 2-time masks with a maximum time mask(T = 50). 
Two convolution sub-sampling layers with kernel size $3\times3$ and stride 2 are used as the frontend. A stack of 4 heads SChunk-Transformer or SChunk-Conformer layers (12 by default) is used as the encoder.
We use a CTC decoder and an Attention decoder of 6 transformer layers with 4 heads. 
The attention dimension is 256 and the feed forward dimension is 2048. 
Accumulating grad is used to stabilize training which updates every 4 steps. Attention dropout, feed forward dropout, and label smoothing regularization are applied in each encoder and decoder layer to prevent over-fitting.
We use the Adam optimizer with the peak learning rate of 0.002  and transformer schedule to train these models for 80 epochs (batch size and warm-up steps are decided based on the memory usage of a model, set to 40 and 25000 by default). And get the final model by averaging the top 20 best models with the lowest loss on the dev set in the training stage.

\subsection{Baseline Systems}
\subsubsection{Chunk-Transformer}
We take the Chunk-Transformer and Chunk-Conformer, which we implemented using WeNet, as the first baseline models. The only difference between them and the proposed models is whether the shifted chunk mechanism is used or not.

\subsubsection{U2}
We take U2\cite{zhang2020unified_arxiv},  a built-in solution in WeNet, as a strong baseline since it's a SOTA model of the time-restricted methods and our models use the same decoder.

\subsection{Ablation Studies}

\begin{table}[t]
\centering
\setlength{\tabcolsep}{4mm}
\caption{ Comparisons with different chunk size (CER$\%$) }
\label{tab:chunk_size}
\begin{tabular}{|c|c|c|c|c|}
\hline
\multirow{2}{*}{Model Architecture} & \multicolumn{4}{c|}{\# Chunk Size} \\ \cline{2-5} 
  &4 &8 & 16 & 32 \\
\hline
Chunk-Transformer &$31.30$ &$18.86$ &$11.80$ &$7.66$\\
Chunk-Conformer &$6.55$ &$6.33$ &$6.09$ &$5.90$\\
\hline
SChunk-Transformer &$7.76$ &$6.68$ &$6.43$ &$5.92$\\
SChunk-Conformer &$6.74$ &$6.21$ &$5.77$ &$5.64$\\
\hline
\end{tabular}
\end{table}

\subsubsection{Chunk Size}
First, we explore how chunk size affects performance. As shown in Table~\ref{tab:chunk_size}, 
we can see that better CERs can be achieved as the chunk size gets larger for both SChunk-Transformer and SChunk-Conformer.
This implies large chunk size is beneficial to capture more global context.
However, we need to balance the accuracy and latency and set the size to 16 for the following experiments.

\subsubsection{\#Encoder Layers}
We also investigate using more
encoder layers to allow sufficient global context capturing.
The results are shown in Table~\ref{tab:layer_number}, the SChunk-Transformer achieves the best
CER with 16 layers, while SChunk-Conformer achieves the best CER using
12 layers. We conjecture this is because the complicated encoder is easier to overfit. 
We set the encoder layer to 12 by default to make our models have similar parameters to others.

\begin{table}[t]
\centering
\setlength{\tabcolsep}{4mm}
\caption{ Comparisons with different number of encoder layers (CER$\%$) }
\label{tab:layer_number}
\begin{tabular}{|c|c|c|c|c|}
\hline
\multirow{2}{*}{Model Architecture} & \multicolumn{4}{c|}{\# Encoder Layers} \\ \cline{2-5} 
  &12 &14 & 16 & 18 \\
\hline
SChunk-Transformer &$6.43$ &$6.25$ &$6.02$ &$6.12$\\
SChunk-Conformer &$5.77$ &$5.81$ &$5.98$ &$6.72$\\
\hline
\end{tabular}
\end{table}

\begin{figure}[t]
  \centering
  \includegraphics[width=8cm]{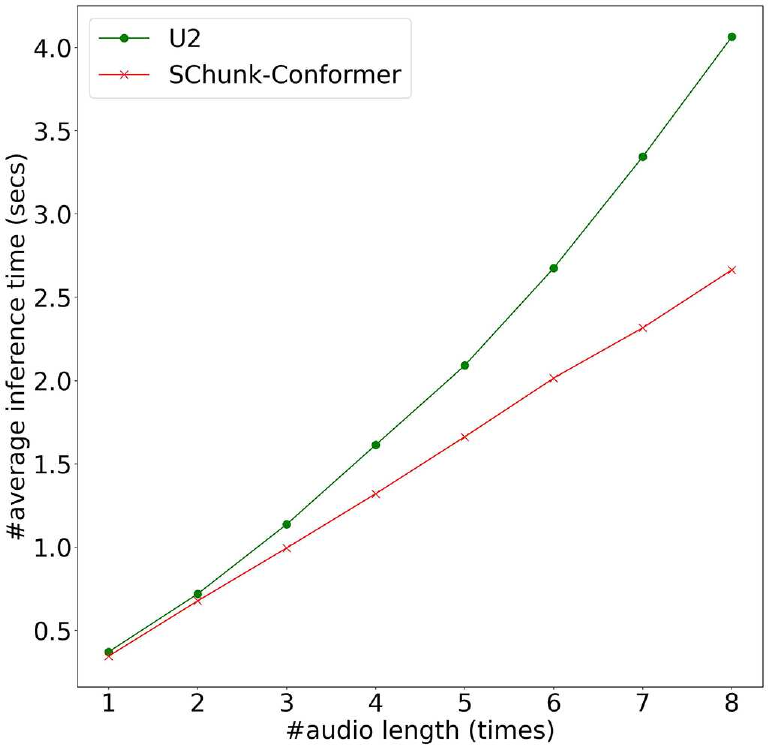}
  \caption{The illustration of inference time cost of U2 and SChunk-Conformer.
We concatenate each audio with itself several times in the test set
of AISHELL-1 to imitate different audio lengths. All the inferences conducted on CPU (Intel(R) Xeon(R) Silver 4210R CPU @ 2.40GHz) with
1-thread, the y-axis indicates the average inference time of 7176 audios.}
  \label{fig:C_U2}
\end{figure}

\subsection{Comparisons with Baseline Systems}
\subsubsection{Chunk-Transformer:}
As shown in Table~\ref{tab:chunk_size}, the CER of Chunk-Conformer is significantly improved compared with Chunk-Transformer, the reason is attributed to the use of CNN to capture sequential history information. With the shifted chunk mechanism, our SChunk-Transformer can
also significantly improve the CER of Chunk-Transformer, which verifies the proposed mechanism can help enhance the ability to model global context. The comparison of SChunk-Conformer and Chunk-Conformer confirms the phenomenon with an exception when the chunk size is 4. This may be because the shifted chunks with attention mask cannot use the whole chunk to model will bring a negative impact in the case of extremely small chunk size.

\begin{table}[t]
\centering
\setlength{\tabcolsep}{3mm}
\caption{ Comparisons with U2. 
max. batch (\#) is the maximum batch size each model can support on two RTX 3090 GPUs,
training time is the total time cost of models trained for 80 epochs with each maximum bath size. }
\label{tab:baseline}
\begin{tabular}{|c|c|c|c|}
\hline
Model Architecture &max. batch (\#)  &trn. time (h) & CER (\%)\\
\hline
U2 (static) \cite{zhang2020unified_arxiv}  &$48$  &$29.13$ &$5.55$\\
U2 (dynamic) \cite{zhang2020unified_arxiv}   &$48$ &$30.56$ &$5.42$\\
SChunk-Conformer  &$60$ &$21.58$ &$5.77$\\
\hline
\end{tabular}
\end{table}

\begin{table}[t]
\centering
\caption{ Comparisons with other streaming solutions (CER\%), $\natural$,$\dag$, and $\ddag$ indicate the solution is a time-restricted method, a chunk-wise method and a memory-based method, respectively.
Following\cite{zhang2020unified_arxiv}, the latency is
defined as the chunk size plus the right context (if any).
$\triangle$ is the additional latency introduced by rescoring. }
\label{tab:other}
\begin{tabular}{|c|c|c|c|c|c|}
\hline
Model Architecture  &Type &Time Complexity &Latency(ms) &LM &CER(\%)\\
\hline
     Sync-Transformer\cite{Tian2020_icassp} &$\dag$ &linear &$400$ & &$8.91$\\
     SCAMA\cite{Zhang2020_interspeech} &$\ddag$ &linear &$600$ & &$7.39$ \\
     MMA-narrow\cite{Inaguma12020_interspeech}  &$\ddag$ &linear &$960$ & &$7.50$ \\
     MMA-wide\cite{Inaguma12020_interspeech} &$\ddag$ &linear &$1920$ & &$6.60$ \\
     HS-DACS \cite{Li2021_head_icassp} &$\dag$ &linear &$1280$ & &$6.80$ \\
     \textbf{SChunk-Transformer(ours)} &$\dag$ &linear & 640+$\triangle$ & &\textbf{6.43} \\
\hline
     U2++ (U2+BiDecoding)\cite{wu2021u2++_arxiv} &$\natural$ &quadratic &$640$+$\triangle$ & &$5.05$ \\
     WNARS(w/ rescoring)\cite{WNARS_arxiv} &$\natural$ &quadratic &$640$+$\triangle$ &$\surd$ &$5.22$ \\
     CUSIDE\cite{CUSIDE_arxiv}  &$\natural$ &quadratic &$400$+$2$ & &$5.47$ \\
     CUSIDE(w/NNLM rescoring)\cite{CUSIDE_arxiv}  &$\natural$ &quadratic &$400$+$2$ &$\surd$ &$4.79$ \\
     \textbf{SChunk-Conformer(ours)} &$\dag$ &linear &$640$+$\triangle$ & &\textbf{5.77} \\
\hline
\end{tabular}
\end{table}

\subsubsection{U2:}
As a strong baseline, U2 can achieve slightly better CER compared with our SChunk-Conformer, see Table~\ref{tab:baseline}. This indicates that the time-restricted methods using full context are beneficial to get better accuracy. However, the performance gap between the SChunk-Conformer and U2 (static, train using static chunk size\cite{zhang2020unified_arxiv}) is quite narrow,
with only 0.22 absolute CER drop. On the other hand, our
models can use a much larger batch size to train, maxium batch size is 60 for SChunk-Conformer while 48 for U2, which can
significantly reduce the training time as shown in Table~\ref{tab:baseline}.
And the average inference time of SChunk-Conformer is linear to the audio length while quadratic for U2, see Fig.~\ref{fig:C_U2}, which is important to control system latency for streaming ASR. All in all, compared with
U2, our models not only can achieve competitive CER, but
also can train and infer more efficiently.

\subsection{Comparisons with Other Streaming Solutions}
Table~\ref{tab:other} lists several recently published Transformer based streaming solutions.
We can see that the SChunk-Transformer can surpass all the chunk-wise or memory-based models,
with 0.37 and 0.17 absolute CER improvement compared with HS-DACS\cite{Li2021_head_icassp} and MMA\cite{Inaguma12020_interspeech}, respectively.
Compared with the other time-restricted models, which use sophisticated techniques (for example, language model (LM) rescoring) to further boost performance compared with U2\cite{zhang2020unified_arxiv},it's not surprise that our SChunk-Conformer fails to achieve superior CER as a chunk-wise model.
However, either U2 or other advanced time-restricted models all have qudratic complexity, in contrast SChunk-Conformer can train and infer more efficiently which is crucial for streaming ASR.

Compared with the other advanced time-restricted models\cite{wu2021u2++_arxiv},\cite{WNARS_arxiv},\cite{CUSIDE_arxiv}, which use sophisticated techniques to further boost performance compared with \cite{zhang2020unified_arxiv}, it’s no surprise that our SChunk-Conformer fails to achieve superior CER as a chunk-wise model. However, either U2 or other advanced time-restricted models all have quadratic time and memory complexities, in contrast, Schunk-Conformer has linear complexity and can train and infer more efficiently which is crucial for streaming ASR.

\section{Discussion}
Our work shows a way to build a single streaming E2E ASR model to achieve the benefits of linear complexity, global context modeling, and parallel trainable concurrently.
Despite the time-restrict models can achieve slightly better CERs, they cannot ensure a low latency in theory makes them impractical for scenarios with long audios.
As the shifted chunk based models can achieve competitive CERs while be insensitive to audio length, they may have a great potential in commercial systems.

\section{Conclusions}
We introduce a shifted chunk mechanism for chunk-wise
Transformer and Conformer models. This mechanism can significantly enhance the modeling power by allowing local self-attention to capture global context across chunks while keeping linear complexity and parallel trainable. Experimental results on AISHELL-1 show that both SChunk-Transformer and SChunk-Conformer can significantly outperform Chunk-Transformer and Chunk-Conformer, respectively. And, the SChunk-Transformer can surpass the SOTA models of both chunk-wise methods and memory-based methods. Compared with the time-restricted methods, our SChunk-Conformer can achieve competitive CER while being able to train and infer more efficiently. In the future, we plan to pay more attention to exploring effective cross-chunk self-attention modeling methods to further improve the performance of streaming ASR.


\end{document}